\documentclass[num-refs]{wiley-article}
\usepackage{siunitx}
\papertype{Original Article}
\paperfield{Journal Section}

\newcommand{\cfps}{CdFeP$_{2}$Se$_{6}$}
\newcommand{\mmpx}{MM$^{'}$P$_{2}$X$_{6}$}

\author[1\authfn{1}]{Manish Kothakonda}
\author[2\authfn{1}]{Yanglin Zhu}
\author[2]{Yingdong Guan}
\author[3]{Jingyang He}
\author[1]{Jamin Kidd}
\author[1]{Ruiqi Zhang}
\author[1]{Jinliang Ning}
\author[3]{Venkatraman Gopalan}
\author[4]{Weiwei Xie}
\author[2]{Zhiqiang Mao}
\author[1]{Jianwei Sun}

\contrib[\authfn{1}]{Equally contributing authors.}

\affil[1]{Physics and Engineering Physics, Tulane University, New Orleans, Louisiana, 70118, USA}
\affil[2]{Department of Physics, The Pennsylvania State University, University Park, Pennsylvania, 16802, USA}
\affil[3]{Department of materials science and engineering, The Pennsylvania State University, University Park, Pennsylvania, 16802, USA}
\affil[4]{Department of Chemistry and Chemical Biology, Rutgers University, Piscataway, New Jersey, 08854, USA}
\corraddress{Jianwei Sun Ph.D., Physics and Engineering Physics, Tulane University, New Orleans, Louisiana, 70118, USA}
\corremail{jsun@tulane.edu}

\title{High-throughput screening assisted discovery of a stable layered anti-ferromagnetic semiconductor: \cfps}

\begin{document}

\begin{frontmatter}
\maketitle

\begin{abstract}
Recent advances in two-dimensional (2D) magnetism have heightened interest in layered magnetic materials due to their potential for spintronics. In particular, layered semiconducting antiferromagnets exhibit intriguing low-dimensional semiconducting behavior with both charge and spin as carrier controls. However, synthesis of these compounds is challenging and remains rare. Here, we conducted first-principles based high-throughput search to screen potentially stable mixed metal phosphorous trichalcogenides (\mmpx, where M and M$^{'}$ are transition metals and X is a chalcogenide) that have a wide range of tunable bandgaps and interesting magnetic properties. Among the potential candidates, we successfully synthesized a stable semiconducting layered magnetic material, \cfps, that exhibits a short-range antiferromagnetic order at $T\textsubscript{N}$ = 21 K with an indirect band gap of 2.23 eV. Our work suggests that high-throughput screening assisted synthesis be an effective method for layered magnetic materials discovery. 

\end{abstract}
\end{frontmatter}

\section{Introduction}

The emergence of two-dimensional (2D) magnetic semiconductors has attracted massive attention, owing to  possible new phenomena arising from 2D magnetism and the
promising potential for spintronic applications \cite{huang2017layer,gong2017discovery,cai2019atomically,long2020persistence}. In particular, antiferromagnetic semiconducting 2D materials have drawn great interest due to their novel physical properties, such as the absence of stray fields, the transmission of spin currents, high intrinsic precision frequency ($\sim$ THz), and high stability under magnetic fields \cite{baltz2018antiferromagnetic,hahn2014conduction, wang2014antiferromagnonic}.  Hitherto, only a handfull of layered antiferromagnetic semiconductors have been predicted and studied, such as CrCl$_3$\cite{mcguire2017magnetic}, double-layer CrI$_3$\cite{jiang2018electric}, FePS$_3$\cite{lee2016ising}, CrOCl \cite{zhang2019magnetism}, H$_x$CrS$_2$\cite{song2019soft}, and Nb$_3$Cl$_8$\cite{haraguchi2017magnetic}. 

\begin{figure*}[htp]
    \begin{center}
    {\includegraphics[width=1\textwidth]{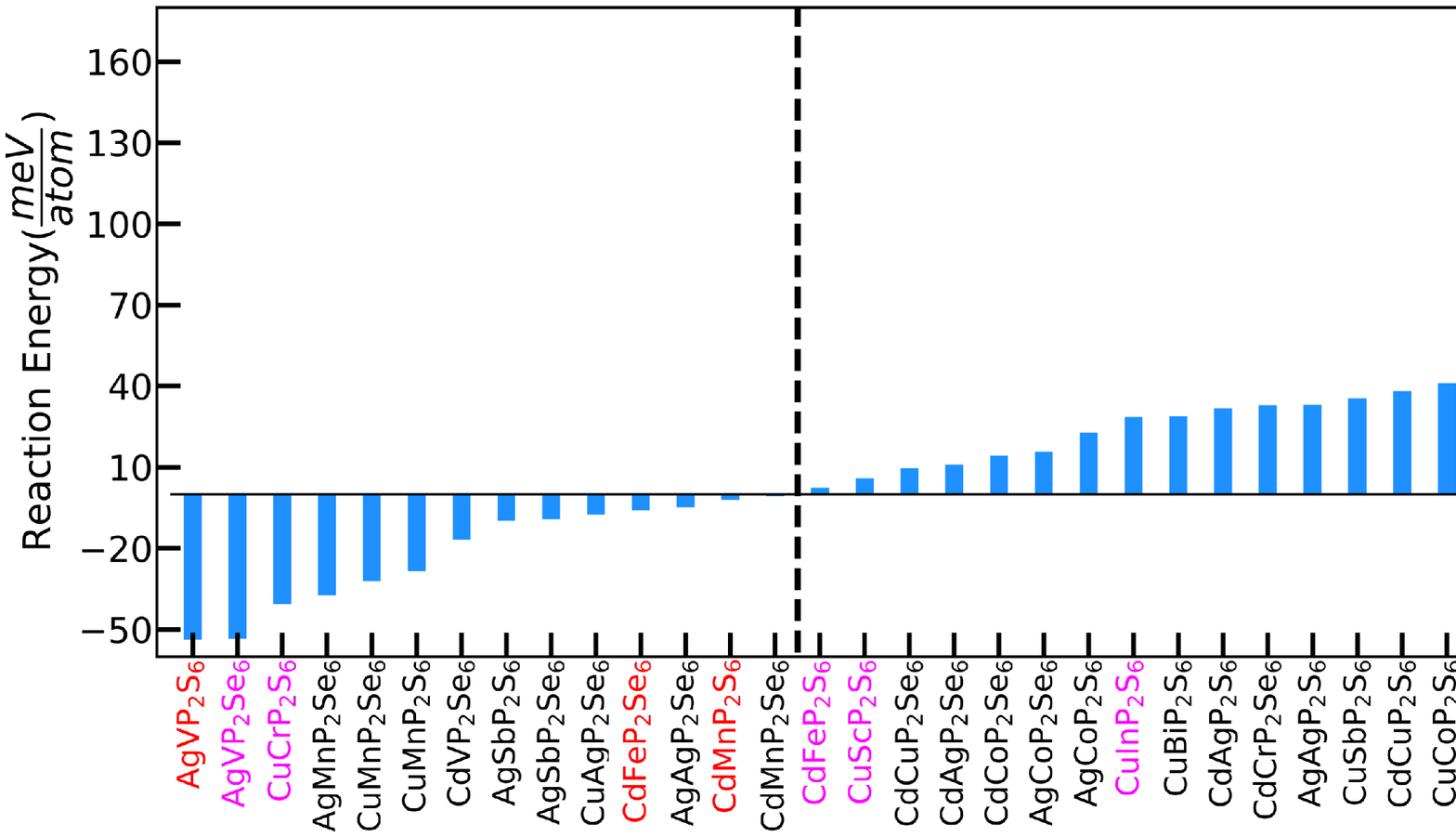}}
    \end{center}
    \caption{High-throughput calculations of reaction energies of 43 \mmpx compositions using DFT with the r$^2$SCAN+D3 functional. The vertical dashed line separates 14 theoretically stable compounds to the left and 29 theoretically unstable compounds to the right. The compositions marked in red were experimentally synthesized in this work. The magenta marked compounds have been previously synthesized experimentally by Refs.  \cite{peng2020quaternary,lee1988general,maisonneuve1995room,lai2019two,belianinov2015cuinp2s6,susner2017metal}.}
    \label{fig:HT_RFE}
\end{figure*}

 The mixed metal phosphorous trichalcogenides, MM$^{'}$P$_2$X$_6$ (M and M$^{'}$ are transition metals, e.g., Cd, V, Mn, Fe, Co, Ni, or Zn, and X is a chalcogenide, i.e., S or Se), are a family of layered materials which have been shown to host stable intrinsic antiferomagnetism (AFM), even at mono- and few-layer thicknesses \cite{lee2016ising}.
 The band gaps of the MM$^{'}$P$_2$X$_6$ materials range from $\sim$ 0.8 eV to $\sim$ 3.5 eV and can be tuned by modifying the metal elements \cite{samal2021two}. Therefore, MM$^{'}$P$_2$X$_6$ could easily be tuned with diverse properties by proper selections of the transition metal and X elements, making it an interesting platform for fundamental science and practical applications. 
 
 In general, novel 2D materials discovery and synthesis is a tedious and expensive trial-and-error experimental process. Only a few of \mmpx compounds have been synthesized in the lab, with thermodynamic stability cited as a major obstacle \cite{peng2021controlling}. 
Given the potential impact of the \mmpx compounds, alternative paradigms are necessary to accelerate the synthesis process. Computations were traditionally used to provide retrospective insights on experimentally synthesized materials and their properties \cite{basnet2022controlling}. For example, Ning et al. \cite{ning2020subtle} showed that first-principles-based free energy calculations predicted the thermodynamic stability of MnBi$_2$Te$_4$ compound and explained the difficulty of synthesizing it. Recent years however witness the predictive power of first-principles methods and the rise of computationally guided materials discovery \cite{furness2020accurate, pan2018data,eng2022theory,torelli2020high}. Here, we conducted a high-throughput search to screen potentially stable \mmpx compounds, leading to the synthesis of \cfps, which is a stable anti-ferromagnetic semiconducting layered material.

\section{High-throughput assisted synthesis of C\MakeLowercase{d}F\MakeLowercase{e}P$_2$S\MakeLowercase{e}$_6$, and determinations of structure and stability}

We performed high-throughput computational screening of 43 \mmpx compounds by calculating the reaction energies. All the magnetic elements were set to ferromagnetic (FM) ordering. The total energies were calculated with the C2 crystal structure observed in the previously synthesized \mmpx compounds \cite{burr1993low}. We used phase diagram hull energy analysis libraries from pymatgen \cite{ong2010thermal,ong2008li} to identify the competing phases of each material for reaction energy calculations (see Sec. \ref{computational method} for details). A negative reaction energy means that the quaternary compound is thermodynamically more stable than its competing phases.

Here, we used the chemical decoration approach to screen stable MM'P$_2$S$_6$ compounds, by changing M and M' to possible transition metal elements Cd, V, Mn, Fe, Co, Ni, and Zn, and S and Se for X. We note that some of the resulting compounds could be more unstable with respect to its competing phases. For example, Ag$_2$MnP$_2$S$_6$ has been experimentally synthesized but not AgMnP$_2$S$_6$\cite{van1993temperature}. We also anticipate that there be false predictions of stable compounds or unstable compounds due to computational accuracy.

Fig \ref{fig:HT_RFE} shows that 14 of the 43 compositions have negative reaction energies, indicating potential synthesizability.
Using the horizontal flux method \cite{yan2017flux}, we have tried to synthesize the 14 potential thermodynamically stable candidates, and only successfully synthesized three single crystals namely, AgVP$_2$S$_6$, \cfps, and CdMnP$_2$S$_6$, highlighted in red in Fig. \ref{fig:HT_RFE}. The procedure used to synthesize these crystals is described in Sec. \ref{experimental method}. We note that AgVP$_2$S$_6$ has been synthesized before \cite{lee1986new,sologubenko2003diffusive}.
Among these three materials, only \cfps shows interesting anti-ferromagnetic and semiconducting properties from experiments that are confirmed by first-principles calculations. We focus on \cfps from now on.

 \begin{figure*}[htp]
    \begin{center}
    {\includegraphics[width=0.92\textwidth]{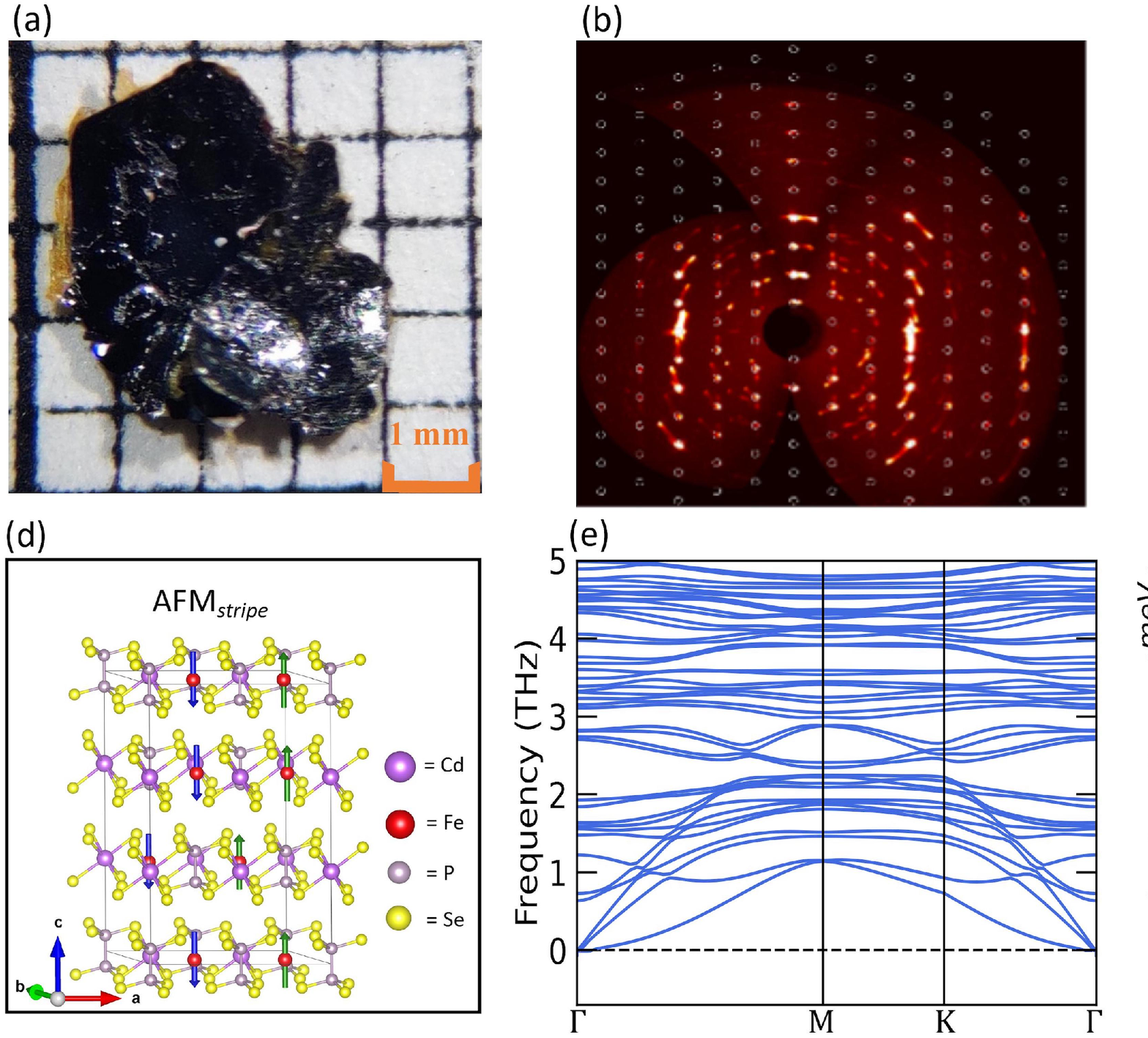}}
    \end{center}
    \caption{(a) Optical image of a \cfps  single crystal. (b) Precision image along (hk0) based on the reflections of \cfps. The intensive reflections match with the observed space group R${\overline{3}}$ and reduced to P3 space group when reduced to the exact stiochiometry of the compound. (c) Atomic force microscopy (AFM) image of a \cfps few-layer flake. Inset: cross-sectional height profile located at the white dashed line. (d) Side view of \cfps in ground state antiferromagnetic stripe magnetic configuration. (e) Vibrational band structure of \cfps.  (f) Reaction free energy(RFE) based on 32CdFeP${_2}$Se${_6}$  → 8Fe${_2}$Se${_4}$ + 8Fe${_2}$P${_4}$ + 16Cd${_2}$P${_2}$Se${_6}$ + Se${_{64}}$. The energy is calculated at two different levels of theory: the 0K DFT reaction energy from r$^2$SCAN+D3 (E$_0$(r$^2$SCAN+D3)) and the  lattice vibrational contribution using the harmonic approximation ((E$_0$(r$^2$SCAN+D3))+ph\_HA).}
    \label{fig:synthesis_strrefin}
\end{figure*}

Figure \ref{fig:synthesis_strrefin}(a) shows an image of a 4×4 mm$^2$ crystal of C\MakeLowercase{d}F\MakeLowercase{e}P$_2$S\MakeLowercase{e}$_6$ obtained from the synthesis. To further refine the crystal structure of the synthesized compound, we performed single crystal X-ray diffraction equipped with Mo radiation. Generally, \mmpx compounds \cite{burr1993low} have the (P$_2$X$_6$)$^{4-}$ anion sublattices that are vertically situated within each layer. The P-P distance within (P$_2$X$_6$)$^{4-}$ is adjusted to accommodate distinct transition metal cations. A larger P-P distance indicates greater thickness of the layer that contains two sub-layers of Se atoms. Transition metal cations, coordinated with six chalcogen anions to form octahedral complexes, are distributed around (P$_2$X$_6$)$^{4-}$ bipyramids in honeycomb configurations. The positions of transition metals then determine the crystal space group.

In our high-throughput screening, the C2 space group \cite{burr1993low} has been used. Our X-ray diffraction measurement, however, indicated that the synthesized \cfps has the space group of R${\overline{3}}$(148), indexed by the reflections as shown in Figure \ref{fig:synthesis_strrefin}(b). The X-ray diffraction measurement can distinguish Fe and Cd atoms. However, the layered compound is too soft to obtain good reflections to generate high-quality refinement. Therefore, the transition metal sites are treated as the same in the determined R${\overline{3}}$(148) space group structure with the Cd:Fe occupation ratio to be 0.43(4): 0.57(4), which indicates the samples have defects. To circumvent the experiments being unable to distinguish Cd and Fe spatially, we assign the Cd and Fe atom to alternative sites, which ultimately reduces the space group of the crystal from R${\overline{3}}$(148) to P3(143).

the X-ray diffraction can distinguish Fe and Cd atoms, that's how we get the approximate atomic ratio of Fe and Cd. However, the layered compound is too soft to obtain good reflections to generate high-quality refinement.


To confirm that the observed R${\overline{3}}$(148), or the simplified P3(143), crystal structure is the thermodynamically most stable structure, we performed DFT calculations of the most commonly seen crystal structures (C2, P1, and P21c) and P3 with possible magnetic configurations. Table \ref{tab:sg_mag_cfps} presents the calculated total energies, lattice constants, and magnetic moments of Fe for the non-magnetic (NM), ferromagnetic (FM), and anti-ferromagnetic with a stripe-like ordering (AFM$_s$) configurations of CdFeP${_2}$Se${_6}$ with different crystal structures. We found that the P3 AFM$_s$ configuration as shown in Figure \ref{fig:synthesis_strrefin}(d) is the most stable among all considered cases, consistent with the X-ray diffraction measurement. It also indicates that the exchange coupling between Fe ions is antiferromagnetic, consistent with the magnetic susceptibility measurement to be discussed later.

Table \ref{tab:sg_mag_cfps} also shows that the optimized lattice parameters of the P3 AFM$_s$ configuration are in good agreement with the experimental values. The in-plane lattice constant of the bulk structure is calculated to be  $a$ = 6.39 \AA, which agrees well with the experimental value of 6.384 \AA. The out-of-plane lattice constant is calculated to be $c = 20.066$ \AA\ (or 6.69 \AA\ of interlayer distance), also agreeing well with the experimental value of 20.011 \AA\ (or 6.67 \AA\ of interlayer distance).

Using the P3 AFM$_s$ structure, we calculate the phonon frequencies, shown in Figure \ref{fig:synthesis_strrefin}(e). There are no imaginary frequencies, indicating that \cfps is dynamically stable. With the calculated phonon frequencies, we include the zero-point vibrational energy and thermal contributions to the reaction free energy with respect to its binary and elemental competing phases Fe${_2}$Se${_4}$, Fe${_2}$P${_4}$, Cd${_2}$P${_2}$Se${_6}$, and Se${_{64}}$. Fig. \ref{fig:synthesis_strrefin}(f) shows the calculated reaction free energy of \cfps, with the dashed line indicating the 0 K ground state energy. The negative reaction free energy confirms the compound to be stable at 0K. The solid line indicates the lattice vibrational contribution under the harmonic approximation. This temperature dependence study shows that with increasing temperature, \cfps is further stabilized. 
\begin{table*}[]
\scriptsize	
\centering
\caption{DFT predictions of total energies with respect to the P3 AFM$_s$ ground state, lattice constants, and magnetic moments of Fe for \cfps. The non-magnetic (NM), ferromagnetic (FM), and antiferromagnetic with a stripe-like ordering (AFM$_s$) configurations  with different crystal structures have been considered.}
\label{tab:sg_mag_cfps}
\begin{tabular}{|l|llll|llll|llll|}
\hline
Sg(No) &
  \multicolumn{4}{c|}{NM} &
  \multicolumn{4}{c|}{FM} &
  \multicolumn{4}{c|}{AFM$_{stripe}$} \\ \cline{2-13} 
 &
  \multicolumn{1}{c|}{Lattice} &
  \multicolumn{1}{c|}{\begin{tabular}[c]{@{}c@{}}Interlayer \\ distance\end{tabular}} &
  \multicolumn{1}{c|}{Energy} &
  M$_{Fe}$ &
  \multicolumn{1}{c|}{Lattice} &
  \multicolumn{1}{c|}{\begin{tabular}[c]{@{}c@{}}Interlayer \\ distance\end{tabular}} &
  \multicolumn{1}{c|}{Energy} &
 M$_{Fe}$ &
  \multicolumn{1}{c|}{Lattice} &
  \multicolumn{1}{c|}{\begin{tabular}[c]{@{}c@{}}Interlayer  \\ distance\end{tabular}} &
  \multicolumn{1}{c|}{Energy} &
  M$_{Fe}$ \\
 &
  \multicolumn{1}{c|}{a} &
  \multicolumn{1}{c|}{d} &
  \multicolumn{1}{c|}{(meV)/f.u.} &
   &
  \multicolumn{1}{c|}{a} &
  \multicolumn{1}{c|}{d} &
  \multicolumn{1}{c|}{(meV)/f.u.} &
   &
  \multicolumn{1}{c|}{a} &
  \multicolumn{1}{c|}{d} &
  \multicolumn{1}{c|}{(meV)/f.u.} &
   \\ \cline{2-13} 
 &
  \multicolumn{1}{l|}{} &
  \multicolumn{1}{l|}{} &
  \multicolumn{1}{l|}{} &
   &
  \multicolumn{1}{l|}{} &
  \multicolumn{1}{l|}{} &
  \multicolumn{1}{l|}{} &
   &
  \multicolumn{1}{l|}{} &
  \multicolumn{1}{l|}{} &
  \multicolumn{1}{l|}{} &
   \\
C2(5) &
  \multicolumn{1}{c|}{6.28} &
  \multicolumn{1}{c|}{6.60} &
  \multicolumn{1}{c|}{777.5} &
  \multicolumn{1}{c|}{0} &
  \multicolumn{1}{c|}{6.37} &
  \multicolumn{1}{c|}{6.65} &
  \multicolumn{1}{c|}{10.8} &
  3.44 &
  \multicolumn{1}{c|}{6.39} &
  \multicolumn{1}{c|}{6.75} &
  \multicolumn{1}{c|}{10.2} &
  3.44 \\
 &
  \multicolumn{1}{l|}{} &
  \multicolumn{1}{l|}{} &
  \multicolumn{1}{l|}{} &
   &
  \multicolumn{1}{l|}{} &
  \multicolumn{1}{l|}{} &
  \multicolumn{1}{l|}{} &
   &
  \multicolumn{1}{l|}{} &
  \multicolumn{1}{l|}{} &
  \multicolumn{1}{l|}{} &
   \\
P1(1) &
  \multicolumn{1}{c|}{6.27} &
  \multicolumn{1}{c|}{6.59} &
  \multicolumn{1}{c|}{777.5} &
  \multicolumn{1}{c|}{0} &
  \multicolumn{1}{c|}{6.35} &
  \multicolumn{1}{c|}{6.70} &
  \multicolumn{1}{c|}{10.9} &
  3.42 &
  \multicolumn{1}{c|}{6.35} &
  \multicolumn{1}{c|}{6.73} &
  \multicolumn{1}{c|}{18} &
  3.42 \\
 &
  \multicolumn{1}{l|}{} &
  \multicolumn{1}{l|}{} &
  \multicolumn{1}{l|}{} &
   &
  \multicolumn{1}{l|}{} &
  \multicolumn{1}{l|}{} &
  \multicolumn{1}{l|}{} &
   &
  \multicolumn{1}{l|}{} &
  \multicolumn{1}{l|}{} &
  \multicolumn{1}{l|}{} &
   \\
P21c(14) &
  \multicolumn{1}{c|}{6.28} &
  \multicolumn{1}{c|}{6.62} &
  \multicolumn{1}{c|}{782.3} &
  \multicolumn{1}{c|}{0} &
  \multicolumn{1}{c|}{6.36} &
  \multicolumn{1}{c|}{6.64} &
  \multicolumn{1}{c|}{20} &
  3.42 &
  \multicolumn{1}{c|}{6.36} &
  \multicolumn{1}{c|}{6.70} &
  \multicolumn{1}{c|}{20} &
  3.42 \\
 &
  \multicolumn{1}{l|}{} &
  \multicolumn{1}{l|}{} &
  \multicolumn{1}{l|}{} &
   &
  \multicolumn{1}{l|}{} &
  \multicolumn{1}{l|}{} &
  \multicolumn{1}{l|}{} &
   &
  \multicolumn{1}{l|}{} &
  \multicolumn{1}{l|}{} &
  \multicolumn{1}{l|}{} &
   \\
P3(143) &
  \multicolumn{1}{c|}{6.29} &
  \multicolumn{1}{c|}{6.61} &
  \multicolumn{1}{c|}{809.2} &
  \multicolumn{1}{c|}{0} &
  \multicolumn{1}{c|}{6.38} &
  \multicolumn{1}{c|}{6.67} &
  \multicolumn{1}{c|}{0.9} &
  3.45 &
  \multicolumn{1}{c|}{6.39} &
  \multicolumn{1}{c|}{6.69} &
  \multicolumn{1}{c|}{0} &
  3.42 \\ \hline
\end{tabular}
\end{table*}

We note that the DFT calculations shown above were performed for the ideal P3 crystal without considering the defects discussed earlier. The defect effects on the stability and properties of \cfps are genuinely interesting. Unfortunately, realistic modeling of the Cd and Fe mixing indicated by our X-ray diffraction would require a large supercell and be computationally prohibitive.

Further complication comes from the difficulty in determining the defects in the synthesized samples. Using an energy dispersive X-ray spectrometer (EDS), the chemical compositions of the grown crystals were determined to be Cd$_{1.06}$Fe$_{0.84}$P$_{2.06}$Se$_6$, while the X-ray diffraction indicated the Cd and Fe atoms mix on one site with the Cd:Fe occupation ratio of 0.43(4):0.57(4). We note XRD and EDS are different techniques. Single crystal XRD can only measure a small piece of crystal, ~20$\mu$m$\times$20$\mu$m$\times$20$\mu$m, while EDS can measure a much larger area, for example, 1mm$\times$1mm$\times$1mm. So, the difference in Cd:Fe chemical ratio between XRD and SEM-EDS is reasonable, while it is certain that the samples have defects.

To demonstrate \cfps to be a potential candidate for 2D spintronic devices, we performed mechanical exfoliation and obtained 2D atomic crystals, as shown in Fig \ref{fig:synthesis_strrefin}(c). Our preliminary trials have obtained the 4-layer ($\sim$2.7 nm) \cfps nanoflake. The thinnest part of the flake is close to the monolayer ($\sim$0.68 nm). These results indicate that this compound holds great potential to realize the 2D AFM state in its monolayer.

\section{Properties of C\MakeLowercase{d}F\MakeLowercase{e}P$_2$S\MakeLowercase{e}$_6$}

\subsection{Optical properties}
Our optical measurements and DFT calculations demonstrated that the synthesized \cfps is a semiconductor with an indirect bandgap. The optical bandgap of \cfps was measured on its (0001) surface using spectroscopic ellipsometry. The complex ordinary optical constant  $\tilde{n}_{o}=n+i k$ (shown in Figure \ref{fig:ordinary_optical_constants}) was extracted by fitting the ellipsometry data to five Tauc-Lorentz oscillators (\ref{tab:tauc_lorentz_oss}). The absorption coefficient can thus be calculated by 
\begin{equation}
\alpha=\frac{4 \pi k}{\lambda}
\end{equation}
where $\lambda$ is the wavelength. The Tauc plot of  $(\alpha h \nu)^{0.5}$ vs. incident photon energy, $\mathcal{E}$, is shown in Figure \ref{fig:bandgap_magnetic}(a), revealing an indirect bandgap of 2.23 eV. The feature in the spectral range below 2.23eV is likely due to the defect states. The direct bandgap of \cfps is also extracted by plotting $(\alpha h v)^{2}$  vs. $\mathcal{E}$ as shown in Figure \ref{fig:bandgap_magnetic}(a). Figure \ref{fig:bandgap_magnetic}(b) shows the DFT band structure of the P3 AFM$_s$ configuration for \cfps. As expected ~\cite{yang2016more}, DFT yields a bandgap ($\sim$ 1eV) that is smaller than the experimental indirect bandgap.

\subsection{Magnetic properties}
\begin{figure*}[ht]
 \centering
{\includegraphics[width=0.8\textwidth]{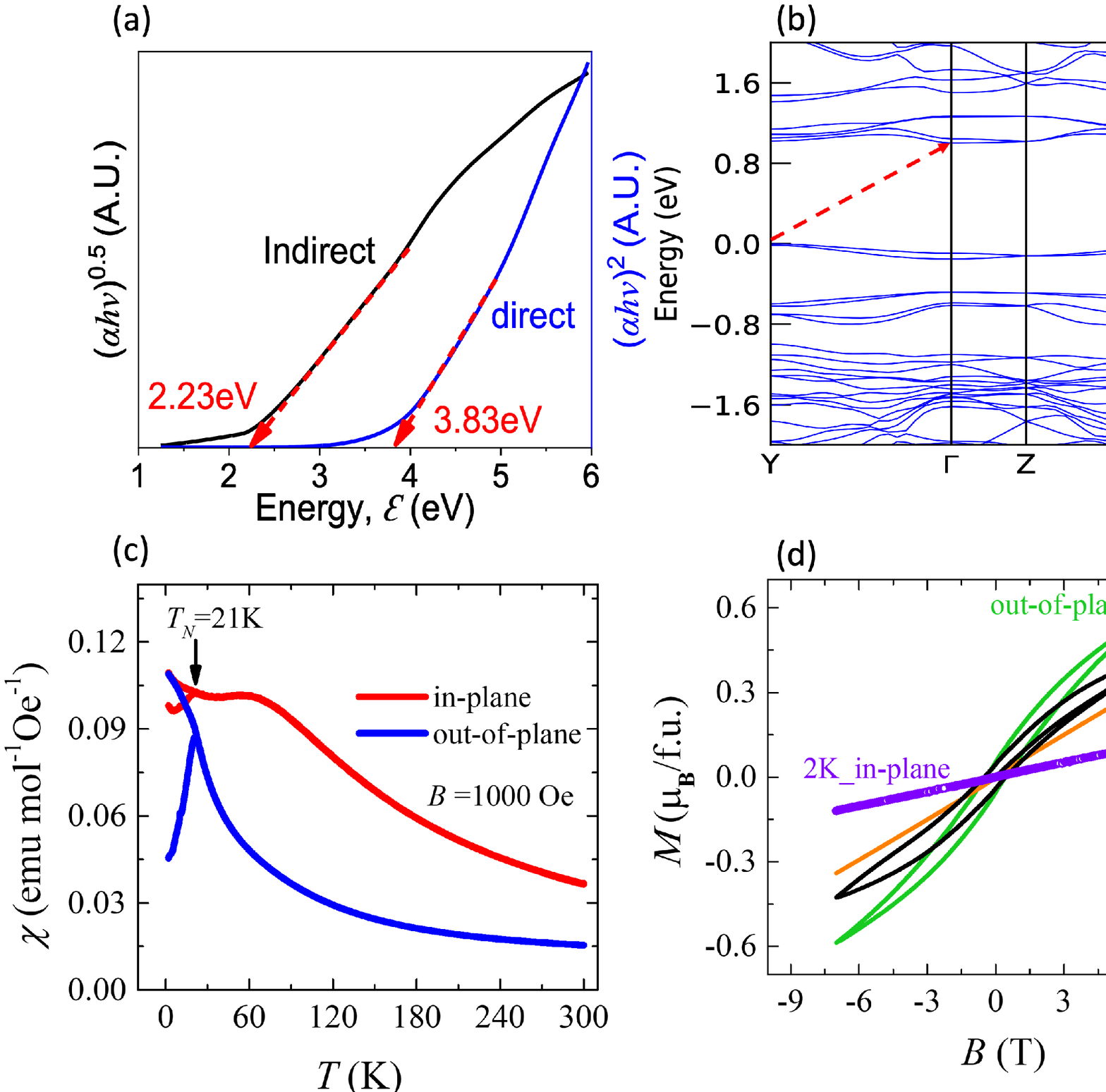}}
\caption{Electronic and magnetic properties of bulk CdFeP${_2}$Se${_6}$. (a) Tauc plot of \cfps, indicating an indirect bandgap of 2.23eV and a direct optical transition of 3.83 eV. (b) Electronic band structure of the ground-state P3 AFM$_{stripe}$ of \cfps. (c) The magnetic susceptibility measured with $B$ applied along the out-of-plane (blue) and in-plane (red) directions. (d) The isothermal magnetization at various temperatures when $B$ is applied along the out-of-plane and in-plane directions.}
\label{fig:bandgap_magnetic}
\end{figure*}

To verify the predicted antiferromagnetic ground state, we performed a systematic magnetization measurement on \cfps single crystals, as shown in Figure \ref{fig:bandgap_magnetic}(c-d). Fig. \ref{fig:bandgap_magnetic}(c) displays the magnetic susceptibility (M-T), and Fig. \ref{fig:bandgap_magnetic}(d) shows the isothermal magnetization (M-H) of \cfps crystals under the magnetic field applied in the $ab$ plane (in-plane) and along the $c$ axis (out-of-plane). 

Fig. \ref{fig:bandgap_magnetic}(c) shows that the zero-field cooling (ZFC) and field-cooling (FC) magnetic susceptibility curves which display a bifurcation at $T$\textsubscript{order} = 21 K when the magnetic field is applied along the $c$ axis. This bifurcation is due to thermomagnetic irreversibility (TMI). Moreover, such irreversibility is also observed when the magnetic field $B$ is aligned in-plane, as shown in the red curves in Fig. \ref{fig:bandgap_magnetic}(c). This likely indicates a short-range magnetic order formed below 21 K. Our heat capacity measurement further confirms the short-range order, as shown in supporting information (SI) Fig. \ref{fig:Temp_dept_spheat}(a). There is no heat capacity anomaly observed around 21K, ruling out the long-range magnetic order in this sample. 
Besides, we noticed a considerable difference between magnetic susceptibility with $B$ aligned in-plane and out-of-plane, indicating the \cfps exhibited strong magnetic anisotropy. This is also revealed by our isothermal magnetization shown in Fig. \ref{fig:bandgap_magnetic}(d). The in-plane magnetization at 2K ($\sim$0.15 \textmu$_B$/f.u.) is smaller than out-of-plane magnetization ($\sim$0.4 \textmu$_B$/f.u.), indicating the spin easy axis is along the out-of-plane direction. The origin of such a strong anisotropy in \cfps could be due to strong spin-orbital coupling introduced by heavy element Se.
Moreover, the susceptibility with $B$ aligned  in-plane show a cross-over transition around 80K. It can be attributed to a strong magnetic anisotropy.  We note such a crossover feature has been reported in other systems with spin-glass and cluster spin-glass systems, such as Ti and Mn-doped Sr$_2$RuO$_4$, which were attributed to a crossover transition from damped inelastic magnetic fluctuations to elastic magnetic order\cite{braden2002incommensurate,ortmann2013competition}. When the inverse magnetic susceptibility was fitted to the Curie--Weiss equation at a higher temperature range ($T$ $>$ 125 K) as shown in SM Fig. \ref{fig:Temp_dept_spheat}(b), we found a negative Curie temperature (-60 K), suggesting antiferromagnetic (AFM) coupling. Such a short-range AFM order is also demonstrated by the isothermal magnetization measurements, as shown in Fig. \ref{fig:bandgap_magnetic}(d). The linear magnetization curves can be observed at 2K when $B$ parallel to the ab plane and at higher temperature ($T$ = 100 K) when $B$ parallel to the c-axis.

Interestingly, the magnetization data for $B$ // c displays a hysteresis at $T< T$\textsubscript{order}.
The presence of a magnetic hysteresis loop  suggests the formation of a static magnetic order. However, as discussed above, such static magnetic order should be in short ranges. It is further supported by the magnetic relaxation measurements shown in Fig. S2c, suggesting glassy behavior in the magnetic state. It is worth pointing out that observation of the magnetic hysteresis does not contradict the short-range AFM order. The  magnetic hysteresis, while more common in ferromagnetic materials, is also often seen in antiferromagnetic materials.\cite{peng2016magnetic,baranov2019magnetic,yan2020type}
Moreover, the magnetic hysteresis  exhibits a relatively large coercive field ($\sim$ 0.45$T$) and a small remnant magnetization (0.05 $\mu_B$/f.u.), which is in accordance with those observed in other magnetic systems with short-range order\cite{monod1979magnetic,li1998evidence,binder1986spin,dho2002reentrant}. Furthermore, at $T$ = 2 K, we noticed that the magnetic moment per unit cell reaches 0.4 $\mu_B$ in 7 $T$, much smaller than the moments of Fe ions. It indicates that the magnetization does not saturate up to 7 $T$, implying that the system may contain magnetic frustration, as suggested by the hexagonal sublattice of Fe ions in \cfps with the P3 crystal structure illustrated in Figure \ref{fig:synthesis_strrefin} (d).

To further understand this observation, we perform magnetic relaxation measurements at various temperatures under ZFC protocol with a magnetic field applied along the $c$ axis, as shown in Fig. \ref{fig:Temp_dept_spheat}(c). The $M$-time curves display the relaxation behavior below $T$\textsubscript{order}. We also found that such magnetic relaxation curves measured at low temperatures ($T$ $\leq$ 20 K) show $\log (t)$ dependence, implying the emergence of a glassy state. Our experimental findings therefore demonstrate that a cluster spin-glass with short-range AFM order emerges at low temperature in \cfps.

Our DFT results for different magnetic configurations (Table 1) show that the AFM$_{s}$ configuration is lower in energy than the FM configuration for the P3 crystal structure. This implies an in-plane AFM coupling that is consistent with the experimental results. To further confirm the in-plane exchange coupling, we investigate the magnetic exchange coupling parameter using the four-states method \cite{xiang2011predicting, xiang2013magnetic}. In our simulations, we only consider the nearest neighbor interaction $J$. As listed in Table \ref{tab:sg_mag_cfps}, the calculated magnetic moment of Fe is $\sim$3.4$\mu_B$. Therefore, the value of spin $S$ in our calculations is set to be 2. Our result shows $J$ = 0.56 meV, suggesting a very weak AFM nearest-neighbor coupling between Fe ions in \cfps, which is consistent with the transition temperature of 21 K. However, the as-grown \cfps exhibits a short-range AFM order, without showing long-range magnetic orders. This is probably due to the strong magnetic frustration dictated by the hexagonal sublattice of Fe ions,
as well as the considerable amount of defects in the synthesized samples revealed by our X-ray diffraction and EDS measurements.

\section{Conclusion}

In conclusion, our high-throughput computational screening of the \mmpx material family led to the discovery of the stable anti-ferromagnetic semiconducting layered material \cfps, demonstrating an effective method to guide 2D material synthesis and accelerate materials discovery. The layered \cfps compound was successfully synthesized and its crystal structure was determined to be R${\overline{3}}$ (148) via the X-ray diffraction measurement. This is confirmed by our DFT calculations, which found the P3 AFM$_{s}$ structure, one closely related to R${\overline{3}}$ (148), to be the ground state among the crystal structures and magnetic configurations considered.
Our magnetic measurements indicated that \cfps hosts a short-range AFM magnetic order below 21 K, consistent with the AFM$_{s}$ magnetic configuration being the ground state and the exchange coupling constant of 0.56 meV from DFT calculations. In addition, DFT is consistent with experiments in predicting an indirect bandgap for \cfps, while the optical measurements found the indirect bandgap to be 2.23 eV. Through microexfoliation, \cfps is exfoliated into a few layers, implying that \cfps could hold excellent potential for realizing the 2D AFM state as a monolayer.

\section{Acknowledgements}
M.K., J.N., and J.S. acknowledge the support of the U.S. Department of Energy (DOE), Office of Science (OS), Basic Energy Sciences (BES), Grant No. DE-SC0014208. J.K. acknowledges the support by the National
Science Foundation Graduate Research Fellowship Program under Grant No. 2139911. Y.L.Z. and Z.Q.M. acknowledges the support from the US DOE under grants DE-SC0019068 for the crystal synthesis and characterization. 
 
\section{Methods}
\subsection{Computational details}
\label{computational method}

All the calculations were performed using the Vienna Ab initio Simulation Package (VASP) version 6.2.1 \cite{kohn1965self,kresse1996efficient}. The projected augmented wave (PAW) method \cite{blochl1994projector,kresse1999ultrasoft} and r$^2$SCAN functional \cite{furness2020accurate} within the meta-generalized gradient approximation (mGGA) were adopted. r$^2$SCAN is a revised version of the strongly constrained and appropriately normed (SCAN) density functional \cite{sun2015strongly} that retains the accuracy of SCAN while improving its numerical stability \cite{furness2020accurate}. SCAN improves over other convectional density functionals for a wide range of properties of correlated materials as reported in previous studies, e.g., for transition metal monoxides~\cite{ZhangPhysRevB2020}, cuprates~\cite{Zhang2020,Furness2018, ChrisLCO2018}, irridates~\cite{ChrisSrIrO42020}, nickelates~\cite{Zhang2021}, and rare earth hexaborides~\cite{ZhangPhysRevB2022}. A cutoff energy of 600 eV was used in structural optimization and self-consistent calculations. The convergence criterion was
10$^{-6}$ eV for the total energy and 0.01 eV/\AA\ for forces during structural optimizations. To account for van der Waals interactions, Grimme's D3 dispersion correction \cite{grimme2010consistent,grimme2011effect} to the total energy was applied throughout the calculations. 

For the high-throughput search studies, we calculated reaction energies (REs) of quaternaries with respect to their respective competing phases. Competing phases of each of the quaternary compounds are determined by analysing phase diagrams attained from pymatgen \cite{ong2010thermal,ong2008li}. For example, in \cfps, with the energy hull analysis of phase diagram from pymatgen, the compounds  8Fe${_2}$Se${_4}$, 8Fe${_2}$P${_4}$,  16Cd${_2}$P${_2}$Se${_6}$, Se${_{64}}$ are determined to be the competing phases. These compounds are then balanced stiochiometrically to form the following reaction:
\begin{equation*}
\footnotesize{
32 \mathrm{CdFeP}_{2} \mathrm{Se}_{6} \rightarrow 8 \mathrm{Fe}_{2} \mathrm{Se}_{4}+8 \mathrm{Fe}_{2} \mathrm{P}_{4}+16 \mathrm{Cd}_{2} \mathrm{P}_{2} \mathrm{Se}_{6} + \mathrm{Se}_{64}}
\end{equation*}
The total energy of each compound in this equation is then used to predict the RE. Negative RE means the quaternary is more stable than its competing compounds, making it thermodynamically stable. All the high-throughput calculations for quaterneries are performed using the C2 phase (initial guess) and FM magnetic ordering.

To assess the thermodynamical stability of \cfps at finite temperatures, we conducted the reaction free energy calculations with lattice vibrations considered at the level of harmonic approximation (HA) \cite{ning2020subtle}. We used the finite displacement method for the phonon calculations within the Phonopy code \cite{phonopy}.

\subsection{Experimental synthesis and characterization} \label{experimental method}

We synthesized \mmpx single crystals by the horizontal flux method \cite{yan2017flux}. The synthesis procedure consists of two parts. First, the stoichiometric M, M', P, and X powder were ground and pressed to a pellet. Then, the pellet was sealed into a quartz tube under vacuum and loaded into a furnace. The furnace was heated to 900$^\circ$C and held for 24 hours for homogeneous melting. Then, the furnace was shut down, and the material was cooled down to room temperature. The preferred pellet was removed from the quartz tube and reground into fine powder. Secondly, this powder and KCl/AlCl$_3$ eutectic flux were mixed with a molar ratio of 1:20 and loaded into a quartz tube. The eutectic flux KCl/AlCl$_3$ refers to the mixture of KCL and ALCl$_3$ powder with a molar ratio of 1:2. The tube was sealed under vacuum and heated in a horizontal double-zone furnace, with the hot end held at 450$^\circ$C and the cold end at 425$^\circ$C for one week. After that, the KCl/AlCl$_3$ flux was washed off by distilled water, and dark brown plate-like crystals with typical size 4×4 mm$^2$ were obtained, as shown in Fig. \ref{fig:synthesis_strrefin}(a).

The crystal structure and lattice parameters of \cfps were determined by single crystal X-ray diffraction. The chemical compositions of the grown crystals were determined using an energy dispersive X-ray spectrometer (EDS), and the measured composition was Cd$_{1.06}$Fe$_{0.84}$P$_{2.06}$Se$_6$. The magnetization properties of samples were measured by a SQUID magnetometer (Quantum Design).
 
\medskip

\bibliography{HT_CFPS.bib}
\clearpage
\begin{figure}
\textbf{Table of Contents}\\
\medskip
  \includegraphics{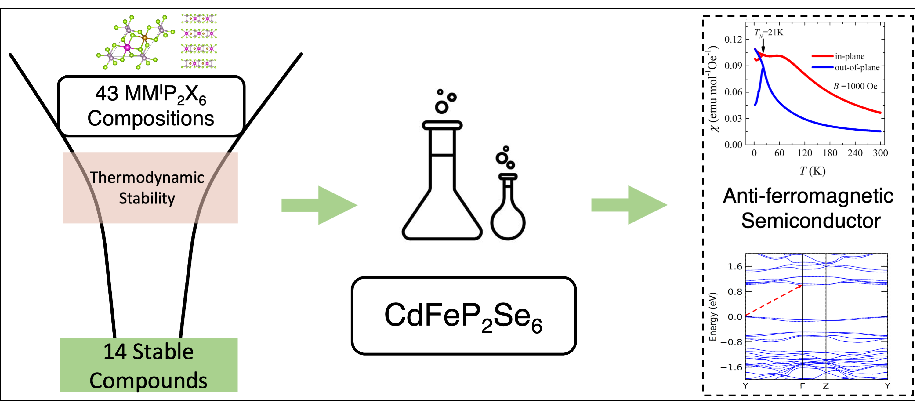}
  \medskip
\end{figure}

\end{document}